\documentstyle[12pt,epsf]{article}

\def\beq{\begin{equation}}
\def\eeq{\end{equation}}
\def\beqa{\begin{eqnarray}}
\def\eeqa{\end{eqnarray}}

\begin{document}
\topskip 2cm 
\begin{titlepage}

\begin{flushright}
ITP-SB-96-30\\
\end{flushright}

\begin{center}
{\large\bf RESUMMATION IN HEAVY QUARK\\
AND JET CROSS SECTIONS\footnote{Presented at {\it Les Rencontres
de Physique de la Vall\'ee d'Aoste}, La Thuile,
Aosta Valley, March 4-9, 1996.}\\
} 
\vspace{2.5cm}
{\large Nikolaos Kidonakis and George Sterman} \\
\vspace{.5cm}
{\sl Insititute for Theoretical Physics\\
State University of New York at Stony Brook\\
Stony Brook, NY 11794-3840, USA}\\
\vspace{2.5cm}
\vfil
\begin{abstract}

We discuss corrections
from the elastic
limit (partonic threshold)
in hadronic hard-scattering cross sections.  We show why these corrections
can be large at all orders in perturbation
theory, and describe their resummation
to arbitrary logarithmic accuracy. 
In particular, we discuss the role of color exchange in the
hard scattering. 
This enables us to generalize the
resummation of the Drell-Yan cross section to QCD reactions.
  As an example, we give the explicit
resummed hard-scattering cross section for heavy-quark production 
through light quark annihilation,
which takes into account next-to-leading
logarithms to all orders.

\end{abstract}

\end{center}

\end{titlepage}

\section{Introduction: Perturbative QCD in Perspective}

Among the motivations for studying perturbative
QCD at high energy are: (i) to test QCD as a quantum field
theory and as a component of the standard model; (ii)
to infer the presence of phenomena
beyond the standard model through deviations from its
predictions and (iii) to understand backgrounds from QCD to signals 
of new particles or nonstandard interactions.    
All of these, but expecially (ii) when the 
deviation from standard model predictions is modest, require us to use
QCD as a precision tool.  Hints of such
deviations that are on everyone's mind
right now are in the running of $\alpha_s$, in $R_b$, and
in very high-$p_T$ jets.  Whether they
persist or not, these examples all suggest the
need to further improve the theory.  
Such improvements will require control
over a class of corrections
associated with what is often called ``partonic threshold",
or more accurately the elastic limit in partonic hard scatttering.  They appear first
at next-to-leading order in hard-scattering cross sections,
and recur in all orders.  This talk will describe 
the nature of these corrections, and report on 
some progress \cite{kidst}
in their resummation to all orders in perturbation theory.

Before going further,
we may distinguish two scenarios for the phenomenological
application of resummed cross sections.  
Corrections due to resummation may turn out to be small, in which case
our confidence in low-order perturbative cross sections should
increase, and our ability to detect new physics
through deviations from QCD predictions
should improve.  Or, they may turn out to be large, 
and may afford
 tests of QCD, and indeed of quantum field theory, in
 a new regime, where all orders of perturbation theory
are relevant.  It is possible that both scenarios apply in 
 different cross sections.
It's a ``win-win" situation.  We begin by reviewing a few facts about the
calculation of hard-scattering cross sections in perturbative QCD.

\section{The Elastic Limit in Hard Inclusive Scattering}

\subsection{Factorized Cross Sections}

We will be interested in inclusive cross sections at 
large momentum transfer through strong interactions.  
In such cross sections we sum over all final states 
that include a particular heavy system $F$, 
which can only be  produced by a short-distance
process in partonic scattering.  Outstanding 
examples of $F$ are 
a top-antitop pair, or a pair of jets at very high
transverse momentum.    
   
\begin{figure}[ht]
\centerline{\epsffile{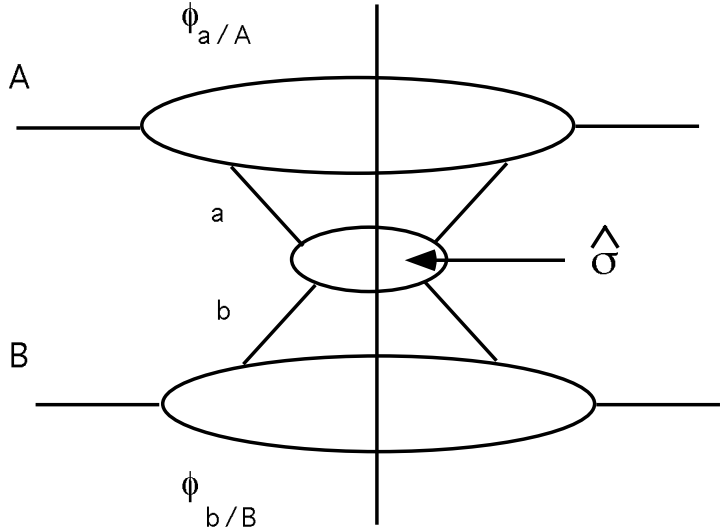}}
\caption{Hard-scattering cross section in cut (unitarity) diagram
notation.}
\label{cutcross}
\end{figure}

We suppose for simplicity that the total mass $Q$ of the system $F$
is of order $S$, the total (hadronic) center of mass energy squared,
and that the rapidity $y$ of the produced system is not large.
Any such cross section can be computed
by combining parton distributions with perturbative
calculations in the factorized expression \cite{fact}
\beqa
{d\sigma_{AB\rightarrow FX}\over dQ^2dy}
&=&
\sum_{ab}
\int_{Q^2/S}^1dz\;
\int\frac{dx_a}{x_a}\frac{dx_b}{x_b}\; \phi_{a/A}(x_a,Q^2) 
\phi_{b/B}(x_b,Q^2) \nonumber \\
&\ & \quad \times \delta\left(z-{Q^2 \over x_ax_bS}\right)\; 
\, 
{\hat \sigma}_{ab\rightarrow FX}\left (z,y,x_a/x_b,
\alpha_s(Q^2)\right)\, ,
\label{basicfact}
\eeqa
which is illustrated in fig.\ \ref{cutcross}.   The $\phi$'s are
usual parton distributions (in some factorization
scheme, like DIS or $\overline{\rm MS}$), and
$\hat{\sigma}$ is a partonic hard-scattering function, which
at lowest order (parton model) is the Born cross section for
$a+b\rightarrow F+X$,
\beq
{\hat\sigma}=\sigma_{\rm Born}+{\alpha_s\over\pi}{\hat\sigma}^{(1)}
+\dots\, .
\label{sigmaborn}
\eeq
$\hat{\sigma}^{(1)}$ is known for many processes, notably
Drell-Yan \cite{dy1loop}, direct photon \cite{dgamma1loop}, 
heavy-quark \cite{heavycalcs}, and jet production \cite{jet1loop}.
We will be interested in the elastic limit (see below), or partonic
threshold of this function.  Two- 
(and sometimes even three-) loop corrections of this sort are
also known in deeply inelastic scattering (DIS) and Drell-Yan 
(DY) cross sections \cite{2loop}.

\subsection{What Threshold?}

The kinematics of the partonic process require that 
$x_ax_bS\ge Q^2$, so that $z\le 1$
in eq.\ (\ref{basicfact}).  At $z=1$,
the partons have just enough energy to produce the observed final
state, with no extra hadronic radiation.  This is what we
shall refer to as the ``elastic limit", or ``partonic threshold".
It is important to distinguish partonic threshold from the usual concept of
a threshold.  In particular, in heavy quark production, we shall
assume that the heavy quarks of mass $M$ are produced with nonzero velocity
$\beta$, and hence with a total invariant mass $Q^2>4M^2$.  Thus,
only for $\beta=0$ does partonic threshold coincide with 
true threshold.  For the Drell-Yan production of 
highly relativistic lepton pairs
with $Q^2\gg 4m_\ell^2$,
partonic threshold still refers to $z=1$, and
is the source of potentially large corrections.

\subsection{Why Large?}

Typical hard-scattering cross sections are distributions 
in the variable $z$ rather
than simply functions of $z$, because they include contributions from
virtual as well as real gluons. 
We are interested in a class of large, positive corrections 
due to such distributions that
occur in all $\sigma^{(n)}$.  Let us explain in what sense they are
``large", and why they are positive to all orders.

At order $\alpha_s^n$, the leading logarithmic distributions in eq.\
(\ref{basicfact}) are of
the form \cite{dy1loop,2loop,oldDY}
\beq
-{\alpha_s^n\over n!}\; \left [ {\ln^{2n+1}\left((1-z)^{-1}\right)\over 1-z} \right ]_+\, ,
\label{lln}
\eeq
whose integral with a smooth function ${\cal F}(z)$ (such as the convolution
of parton distributions in eq.\ (\ref{basicfact})) is
\beqa
-{\alpha_s^n\over n!}\int_0^1dz\; 
{{\cal F}(z)-{\cal F}(1)\over 1-z}\; \ln^{2n+1}\left((1-z)^{-1}\right)
&=&{\alpha_s^n\over n!}\int_0^1dz\; {\cal F}'(1) \ln^{2n+1}\left((1-z)^{-1}\right)+\dots \nonumber \\
&\sim& {\alpha_s^n\over n!}{(2n+1)!}+\dots
\label{factorial}
\eeqa
where we have kept only the first term in the expansion of ${\cal F}(z)$ 
about $z=1$.  It is evident that such terms give, at least
formally, contributions that grow even faster than $n!$ at $n$th order.
If they had alternating signs,
these contributions might add up to a finite number somehow, but
they are all of the
same sign.  

\subsection{Why Positive?}

\begin{figure}[ht]
\centerline{\epsffile{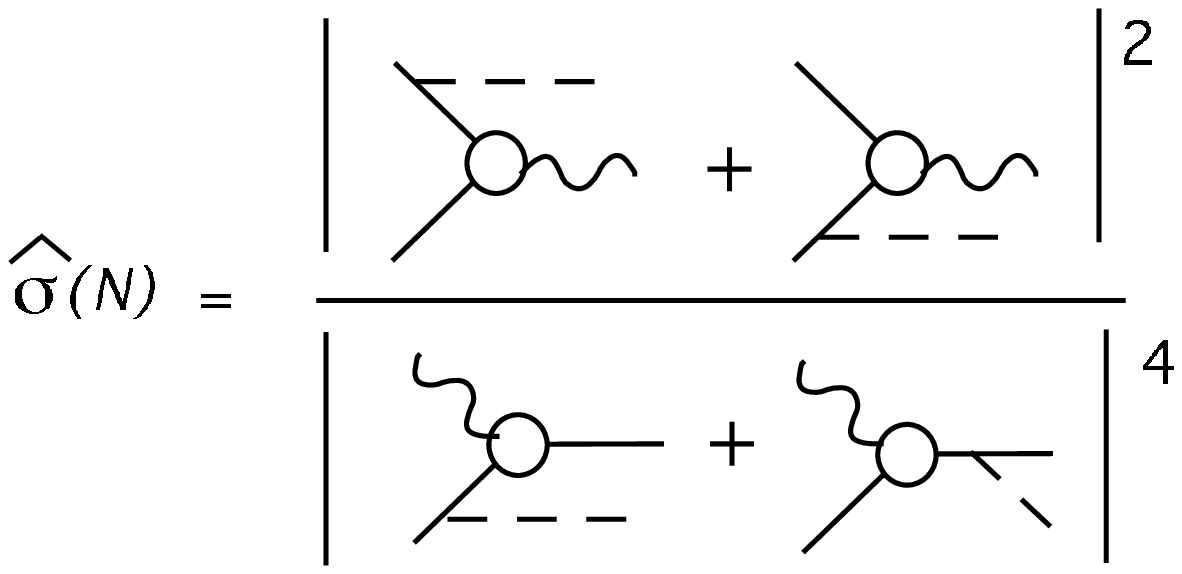}}
\caption{Schematic representation of moments of the
Drell-Yan partonic hard-scattering function.}
\label{mtratio}
\end{figure}

Why are these corrections positive, and hence potentially
dangerous?  Their sign comes directly from the manner in which
hard-scattering cross sections are computed.  
The fully inclusive Drell-Yan cross section $d\sigma/d Q^2$ illustrates the
situation.
The computation of its hard-scattering function
is easiest to understand in terms of moments, because 
eq.\ (\ref{basicfact}) factors into simple products 
of functions under
moments with respect to $\tau=Q^2/S$,
\beq
\hat{\sigma}(N)\equiv 
{1\over\left[\int_0^1dx\; x^{N-1}\; \phi(x)\right]^2}\, 
\int_0^1 d\tau \tau^{N-1}{d\sigma_{\rm DY}(\tau)\over dQ^2}\, .
\eeq
  Neglecting parton labels, the
moment $\hat{\sigma}(N)$ is the ratio of moments of the cross section
to the product of moments of parton distributions.  Because $\hat{\sigma}$
is, by construction, dependent only on short-distance behavior, 
this ratio may be computed in perturbation theory, as illustrated
schematically in fig.\ \ref{mtratio} (for the DIS scheme).   
The numerator is a moment of the perturbative
partonic Drell-Yan cross section,
while the denominator is the product of 
moments of two perturbative parton distributions.  For quark-antiquark
processes, the parton distributions are the same, so the denominator
is the square of squared partonic amplitudes, summed over final states.

At each order both the numerator and denominator in fig.\ \ref{mtratio} have
double-logarithmic terms like eq.\ (\ref{lln}).
All logarithmically-diverent integrals over gluon
transverse momenta cancel in the ratio by  the standard
factorization theorems.  
Before moments, the perturbative Drell-Yan cross sections include 
logarithmic distributions in $1-z$, like eq.\ (\ref{lln}) above,
 while in the deeply inelastic scattering cross section 
the same sort of distributions depend 
on Bjorken $x$ through $1-x$.  After moments, both give double-logarithmic 
$\alpha_s^n\ln^{2n}N$
at $n$th order, with $N$ the moment variable.
These leading
logarithms are the finite remainders of
corrections from $n$ pairs of
 real and virtual gluons
that attach to the scattered quarks in DIS and the annihilating
pair in DY. 
Now, in the denominator, each DIS parton distribution,
which is itself of the form of a cross section, has
both incoming {\it and} outgoing
quarks, while in the numerator, DY involves incoming quarks only.
Simply counting quarks, we discover that
the coefficients of the double logs are $2^n$ times
larger in the denominator than in the numerator at $n$th order.
At the same time, both numerator and denominator have {\it alternating
signs} for their leading logarithms.  The reason for this
may be seen by recalling the relations of $z$ and $x$ to the invariant
mass $W$ of hadrons in the final state for the two cases:
\beqa
{\rm DY}&:& W^2\sim Q^2(1-z)^2\, ,\nonumber \\
{\rm DIS}&:& W^2\sim Q^2(1-x)\, .
\label{wsquared}
\eeqa
The limits $z\rightarrow 1$ and $x\rightarrow 1$ thus both
correspond to nearly elastic scattering: for Drell-Yan, the
annihilation of a quark pair into an electroweak vector boson,
for DIS, the scattering of a quark into a nearly massless jet
of particles.  In a gauge theory like QCD, the alternating-sign
distributions in either cross section sum up to give Sudakov
suppression in the elastic limit.  The perturbative theory simply
will not allow the annihilation or hard-scattering of isolated
colored particles without copious radiation.  
It is not difficult to verify that in these limits, the partonic
cross section is suppressed by a factor that decays faster than
any fixed power of $Q$ \cite{oldDY}.
Indeed, in these
limits, we expect the coherent scattering of hadronic bound states,
whose contributions are normally suppressed by a power of $Q$
compared to incoherent partonic scattering, to dominate.

As a result of the extra
suppression in DIS, due to outgoing
quarks, when the  
hard-scattering function $\hat{\sigma}$, fig.\ \ref{mtratio}
is computed in perturbation theory, the DIS denominator is suppressed
even more than the DY numerator.  Then the ratio actually grows
with moment $N$, from the elastic limit in $z$ space.
This is the source of the terms shown in eq.\ (\ref{lln}) in
$\hat{\sigma}$, and is the reason why they all have the same sign.

\subsection{Resolution?}

For Drell-Yan and other hard-scattering cross sections
the summation of leading singular distributions in $1-z$ is most easily
carried out in moment space, where we find \cite{oldDY},
\beq
\int_0^1 d\tau \tau^N{d\sigma_{\rm DY}(\tau)\over dQ^2}
\sim \exp \left[ + {\alpha_s(Q)\over \pi}C_F\ln^2 N\right ]\, .
\eeq
(An equivalent relation
holds at fixed rapidity \cite{LaenenSterman}.)
The moments, however, require integrals all the way to $\tau=0$,
which implies $S\rightarrow\infty$ at fixed $Q^2$.  In the cross section
itself, which is an inverse moment, the fixed total energy
keeps gluon emission kinematically linked, and we may expect
the inverse transform to be rather better behaved when nonleading
as well as leading contributions are taken into account.
And indeed, recent estimates of the Drell-Yan cross section based
on this approach to resummation give predictions which are
just a few percent (at most) above the exact two-loop results
for fixed target energies.  

There has been considerable (but
perhaps not yet enough) discussion in the literature on just what is the
best way to define and invert a resummed cross section
\cite{CSt}-\cite{catanietal}.  Most of
these issues are well-illustrated by the Drell-Yan case.

\subsection{Resummed Drell-Yan}

The resummed Drell-Yan cross section is the benchmark example for
the resummation of singular distributions.  As above, singular
distributions at $z=1$ translate into logarithms of the
moment variable $N$.  Logarithms of $N$ (to all logarithmic
order, not just leading or next-to-leading logarithm) 
in the moments of the inclusive Drell-Yan cross section 
exponentiate \cite{oldDY},
\beq
\hat{\sigma}_{\rm DY}(N)={d\sigma_{\rm Born}\over dQ^2}\;
e^{C(\alpha_s)+E(N,\alpha_s)}\, ,
\label{dyexp}
\eeq
where $\alpha_s$ stands for $\alpha_s(Q^2)$.  In the exponent,
the function $C$ is known to two loops, while the function $E$,
which organizes all logs of $N$, has the following form in the DIS scheme,
\beqa
E(N,\alpha_s)
&=&
-\int_0^1dx{x^{N-1}-1\over 1-x} \bigg [
\int_0^x{dy\over 1-y}\;
g_1\left(\alpha_s\left[(1-x)(1-y)Q^2\right]\right)
\nonumber\\
&\ & \quad\quad + g_2\left(\alpha_s\left[(1-x)Q^2\right]\right)
\, \bigg]\, .
\label{dyexponent}
\eeqa
The functions $g_1$ and $g_2$ are finite series in $\alpha_s$ \cite{oldDY},
\beqa
g_1(\alpha_s)&=&2C_F
\left ( {\alpha_s\over \pi} + {1\over 2}K\; 
\left({\alpha_s\over \pi}\right)^2\right )+\dots\, ,\nonumber\\
g_2(\alpha_s)&=&-\frac{3}{2}C_F{\alpha_s\over\pi}+\dots\, ,
\label{g12def}
\eeqa
where 
\beq
K=C_A\left({67\over 18}-{\pi^2\over 6}\right ) -{5\over 9}n_f\, .
\eeq
Now eqs.\ (\ref{dyexp}) and (\ref{dyexponent}) resum all
logarithms of $N$ in the sense of an order-by-order
expansion, by reexpanding the running couplings in
terms of $\alpha_s$.  The resummed integrals, however,
are ill-defined for $x\rightarrow 1$, no matter how large
$Q^2$ is, since the one-loop
running coupling $\alpha_s(\mu)
=4\pi/b_1\ln(\mu^2/\Lambda^2)$
diverges at $\mu^2=(1-x)(1-y)Q^2=\Lambda^2$.  
Such a divergence is called an ``infrared renormalon".
The problem of infrared renormalons in
resummed cross sections  \cite{irr} takes its place
alongside the divergence identified above due to
Sudakov logarithms as an object of lively interest.
This is not, however, the subject of this talk.
(It is addressed in Paolo Nason's
contribution to this conference.) Rather, we will report below on
how the resummation of Drell-Yan cross sections
in eqs.\ (\ref{dyexp}) and (\ref{dyexponent}) may be
generalized to include all logarithmic order in 
cross sections like heavy quark or high-$p_T$ jet production, which
are initiated by QCD hard-scattering.  This problem
is distinguished from Drell-Yan by the complications
of final-state radiation, and by color exchange in
the hard-scattering, which is no longer based upon
an electroweak interaction.

\section{Threshold Resummations for Heavy Quarks and Jets}

Over the past few years the one-loop-corrected jet
cross section $d\sigma_{\rm Jet}^{(1)}(p_T)/dp_T$ has
become an almost proverbial success for perturbative
QCD, tracking the data to a few tens of percent
(depending on the parton distributions) while 
it changes over many orders of magnitude.  
Nevertheless, recent experimental results have
afforded a stimulus to study yet higher orders
in QCD cross sections,
particularly in terms of the elastic limit.
Partly, this has come from the desire for resummations in
heavy quark production 
[12-15],
but even more strikingly, from the suggestion of an excess
of events at the very highest jet $p_T$.  

Thus far, resummations for top production have included
leading logarithms in the singular distributions, through the function $g_1$
in eq.\ (\ref{dyexponent}).   It is not difficult
to show that the leading logarithms are the same
for Drell-Yan as for top, and even high-$p_T$ jet,
production.  Resummations for gluons may be included
by simply changing $C_F$ to $C_A$.
For instance,
Ref.\ \cite{jacketal}
on the
one hand, and Ref.\ \cite{bercon} on the other,
start from the same resummation in moment space, but
differ in their treatment of the $n\rightarrow 1-z$
transform.  
Beyond leading logarithms, however, there may be important
differences between the electroweak-induced Drell Yan
cross sections and the QCD-induced top or jet cross sections.
These are due primarily to the presence of final-state
radiation from scattered quarks in the latter case, which
is absent in the former, and to the interplay
of color exchange in the hard scattering with the soft
radiation.  

\subsection{Resummation with Color Exchange}

\begin{figure}[ht]
\centerline{\epsffile{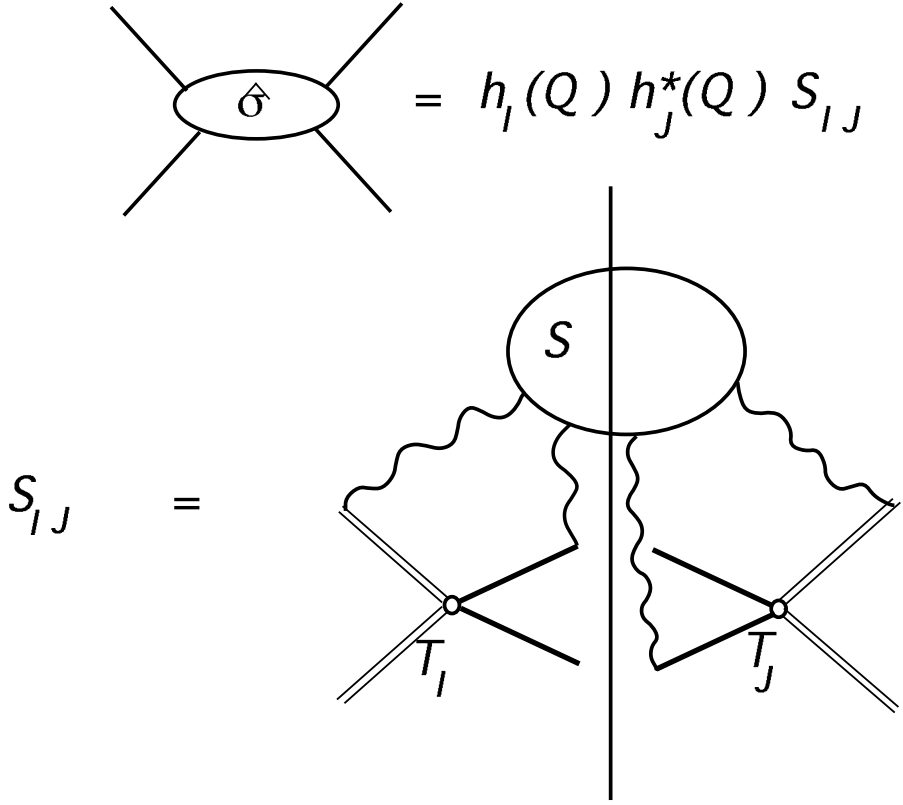}}
\caption{Representation of the factorization
of the hard scattering function $\hat{\sigma}$ 
near the elastic limit.  The second part shows
the soft-gluon matrix $S_{IJ}$ as a
cut diagram for the scattering of incoming ordered
exponentials (double lines - the incoming partons
in the eikonal approximation) to give outgoing
ordered exponentials (bold lines - the outgoing
heavy quarks in the eikonal approximation). For simplicity,
only a few of the possible gluon 
interactions with the ordered exponentials are shown.}
\label{HS}
\end{figure}

An exploration of
the details of resummation in processes based on
a QCD Born cross section, such as heavy quark
and jet production,
requires more time than we can devote here.  As in the case of
Drell-Yan (\ref{dyexp}), resummation is
based first of all on the factorization properties
of the cross section in the neighborhood of the
elastic limit \cite{oldDY}.  The situation is illustrated in
fig.\ \ref{HS}.  
Near the elastic limit, all gluons emitted into the final
state have energies limited by $(1-z)Q\ll Q$.
Correspondingly, gluons with energies of order $Q$
can appear only in virtual states.  Standard
factorization methods may then be used to separate 
the (relatively soft but still perturbative) soft
gluons from the underlying hard scattering.  
We emphasize that this may be done 
order-by-order in perturbation theory, and that
both the factorized hard and soft components
of $\hat{\sigma}$ remain free of soft and 
collinear divergences.  The process of factorization
may be thought of as the construction of an
``effective field theory" \cite{effective} for soft gluons in the
presence of the hard scattering.
 
The hard-scattering function $\hat{\sigma}$ thus
breaks up into a product of 
hard and soft functions. 
In the relevant effective field theory, the incoming
partons that annihilate into the heavy quarks and
the outgoing heavy quarks themselves are represented by
ordered exponentials (Wilson lines) in the directions
of the partons they represent.  The Wilson lines are 
tied together in the amplitude and its complex conjugate
at local vertices, $T_I$ and $T_J$  
in fig.\ \ref{HS}, which describe the flow of color 
between the initial and final state.  
Indices $I$ and $J$
label matrices in color space.
The simplest examples are for the annihilation of light
quarks (color indices $a_1$ and $a_2$)
into heavy quarks (indices $a_3$ and $a_4$), 
\beq
q_{a_1}(p_a)+{\bar q}_{a_2}(p_b) 
\rightarrow Q_{a_4}(p_1) + {\bar Q}_{a_3}(p_2)\, ,
\label{qqQQ}
\eeq
with kinematic invariants,
\beq
t_1=(p_a-p_2)^2-m^2, \quad u_1=(p_b-p_2)^2-m^2, \quad s=(p_a+p_b)^2\, .
\eeq
Here, for instance, we
may choose a basis for the $T$'s that represents
color singlet and octet exchange in the 
$s$-channel,
\beqa
\left( T_1 \right)_{\{a_i\}}&=&\delta_{a_1a_2}\delta_{a_3a_4}\, ,
\nonumber\\
\left( T_2 \right)_{\{a_i\}}&=&
\sum_c \left(T_c^{(F)}\right)_{a_2a_1}\left(T^{(F)}_c\right)_{a_4a_3}\, .
\label{Tdefs}
\eeqa
Other bases, particularly singlet exchange
in the $s$- and $u$- channels, are also interesting.
As in fig.\ \ref{HS}, each choice of effective vertices
leads to a separate soft function $S_{IJ}$, which
depends on $(1-z)Q$ only, rather than $Q$ itself.

Meanwhile, the two
virtual hard-scattering functions, $h_I(Q)$ and $h^*_J(Q)$, 
which contain only virtual corrections and hence depend
on $Q$ only, are labelled by
the same color exchange indices.  As in most factorizations
and constructions of effective field theories, the new
 vertices require renormalization.  Thus
we renormalize the soft functions \cite{BottsSt},
\beq
S_{IJ}^{({\rm un})}\left((1-z)Q\right)=
Z_{II'} Z^*_{JJ'}S_{I'J'}^{({\rm ren})}\left((1-z)Q\right)\, ,
\label{Sren}
\eeq
and the hard functions
\beq
h_I^{({\rm un})}(Q)=Z^{-1}_{IJ}h_J^{({\rm ren})}(Q)\, ,
\label{hren}
\eeq
where the $Z_{KL}$ are cutoff-dependent renormalization  constants.
The renormalization of such composite vertices linking
Wilson lines has been discussed elsewhere, primarily 
in the context of nearly forward scattering for lightlike
ordered exponentials [18-21].
In our case, to make a long story short, 
$1-z$ dependence beyond leading logarithm is determined
by the anomalous dimension matrix corresponding to
this renormalization \cite{BottsSt,GK,KK,CLS}, in a manner analogous to the
way in which the evolution of singlet parton
distributions is controlled by the anomalous dimensions
of light-cone operators in DIS.
  
In general, solutions to the renormalization group equation for $S_{IJ}$
that follows from (\ref{Sren}) are ordered
exponentials \cite{BottsSt,CLS}, as for singlet evolution in DIS. 
At leading logarithm in $S_{IJ}$, which is next-to-leading
logarithm in the overall cross section, however, we can
diagonalize the anomalous dimension, and separate the
evolution of particular linear combinations of composite color
vertices.  This is a generalization of the exponentiation
of logarithms (and infrared
divergences) in the Sudakov form factor \cite{Sudakovff}.

Passing from these general considerations to specific results,
let us give the resummation of singular distributions
at $z=1$ for the partonic process in eq.\ (\ref{qqQQ}) to
next-to-leading logarithm in $1-z$ at all orders
for light quark annihilation into heavy quarks. 
We consider the production of a pair of heavy quarks
with total invariant mass $Q\ge 2M_Q$ at rapidity $y$.
The cross section is, as usual, a convolution 
of hard-scattering functions $\hat{\sigma}_{ab}$ with
parton distributions $\phi_{q/A}$ and $\phi_{{\bar q}/B}$,
as in eq.\ (\ref{basicfact}), with $F=Q{\bar Q}$.
Corresponding to the Drell-Yan result, (\ref{dyexp}), we now have,
to next-to-leading logarithmic accuracy,
\beq
\hat{\sigma}_{q{\bar q}\rightarrow Q{\bar Q}}(N)
=\sum_{IJ}S^{(0)}_{IJ}\; h_I(Q)h^*_J(Q)\;
e^{C'(\alpha_s)+E_{IJ}(N,\alpha_s)}\, ,
\label{qqQQexp}
\eeq
where again $\alpha_s$ stands for $\alpha_s(Q^2)$.  
The function $C'$ is known to one loop only at
this time.  To next-to-leading log,
we need only the lowest-order 
soft functions, $S_{IJ}^{(0)}\sim\delta_{IJ}$.
The function $E_{IJ}$,
which contains the logs of the moment variable
$N$ has a form very similar to the Drell-Yan case, but now
with a dependence on the effective color vertices, through
a third function, $g_3$,
\beqa
E_{IJ}^{(ab)}(N,\alpha_s)
&=&
-\int_0^1dx{x^{N-1}-1\over 1-x} \bigg [
\int_0^x{dy\over 1-y}\;
g_1\left(\alpha_s\left[(1-x)(1-y)Q^2\right]\right)
\nonumber\\
&\ & \quad\quad + g_2\left(\alpha_s\left[(1-x)Q^2\right]\right)
\nonumber\\
&\ & \quad\quad + g_3^{(I)}\left(\alpha_s\left[(1-x)^2Q^2\right]\right)
+ g_3^{(J)}{^*}\left(\alpha_s\left[(1-x)^2Q^2\right]\right)\, \bigg]\, .
\label{qqQQexponent}
\eeqa
As before, the $g_i$, $i=1,2,3$ are finite functions of their
arguments.  In the DIS scheme, $g_1$ and $g_2$ are
given for incoming light quarks by (\ref{g12def}) above.
Dependence on color exchange in the hard
scattering is contained entirely in 
the new functions $g_3^{(I)}$, which may 
(but need not) be defined
to be zero in Drell-Yan, eq.\ (\ref{dyexponent}) \cite{CT2}. 
To determine $g_3^{(I)}$, we
go to a color basis that diagonalizes the
renormalization matrix $Z_{IJ}$ in eqs.\ (\ref{hren})
and (\ref{Sren}).  In this basis,
\begin{equation}
g_3^{(I)}[\alpha_s]=-\lambda_I[\alpha_s]\, ,
\end{equation}
where the eigenfunctions $\lambda_I$ are complex in general, and may depend
on the directions of the incoming and outgoing partons.

We are now ready to give the anomalous dimension matrix
of the effective vertices $T_I$ in fig.\ \ref{HS} for
light to heavy quark annihilation in the 
singlet-octet basis of eq.\ (\ref{Tdefs}) \cite{kidst}:
\begin{eqnarray}
\Gamma_{11}&=&-\frac{\alpha_s}{\pi}C_F(L_{\beta}+1+\pi i),
\nonumber \\
\Gamma_{21}&=&\frac{2\alpha_s}{\pi}
\ln\left(\frac{u_1}{t_1}\right),
\nonumber \\ 
\Gamma_{12}&=&\frac{\alpha_s}{\pi}
\frac{C_F}{C_A} \ln\left(\frac{u_1}{t_1}\right),
\nonumber \\
\Gamma_{22}&=&\frac{\alpha_s}{\pi}\left\{C_F
\left[4\ln\left(\frac{u_1}{t_1}\right)-L_{\beta}-1-\pi i\right]\right.
\nonumber \\ &&
\left.+\frac{C_A}{2}\left[-3\ln\left(\frac{u_1}{t_1}\right)
-\ln\left(\frac{m^2s}{t_1u_1}\right)+L_{\beta}+\pi i \right]\right\}\, .
\label{gammaoneloop}
\end{eqnarray}
Here $L_\beta$ is the vertex function in the eikonal approximation for the 
production of a pair of heavy quarks with center of mass velocity $\beta$,
\beq
L_{\beta}=\frac{1-2m^2/s}{\beta}\left( \ln\frac{1-\beta}{1+\beta}
+i\pi\right), 
\quad \beta=\sqrt{1-4m^2/s}\, .
\eeq
Solving for the eigenvalues, substituting them in eq.\ (\ref{qqQQexp}),
and expanding the result to first order in $\alpha_s$,
we can derive an explicit one-loop expression for the
cross section for heavy quark production through light
quark annihilation.  We have checked that this result is
consistent with the explicit one-loop formulas given 
in \cite{mengetal}.    Here we content ourselves with pointing out
that, unlike leading logarithms, next-to-leading logs depend
on angles, through ratios of kinematic invariants, such as
$u_1/t_1$ and $s/t_1$.   Interestingly, the singlet-octet anomalous dimension matrix
is manifestly diagonal at ninety degrees.

Much the same considerations apply
to jet production, whose resummation requires an
additional factorization of the collinear singularities within 
the jets from the hard
scattering {\it and} soft emission.  In particular, the
anomalous dimension matrix in this case is also 
dependent on the direction of the jets.  The
effect of resummation at next-to-leading
logarithm will therefore in general change the
angular dependence of the cross section, relative
to next-to-leading order.

\section{Conclusion}

The general considerations and the explicit results quoted
above are part of a renewed phenomenology of higher-order
corrections in perturbative QCD.  It remains to be seen
if, and where, resummed next-to-leading logarithmic
corrections like those
quoted above are phenomenologically significant. As we
have discussed above, however, the results will be of interest
whether they are large
or small.

\noindent
\section*{Acknowledgements}

  We wish to thank Jack Smith and
Eric Laenen for useful conversations.
This work was supported in part by the National Science Foundation under
grant PHY9309888.


\begin{thebibliography}{99}

\bibitem{kidst} N.\ Kidonakis and G.\ Sterman, Stony Brook preprint ITP-SB-96-7,
hep-ph/9604234, and in preparation; N.\ Kidonakis, Ph.D.\ thesis,
Stony Brook, 1996.

\bibitem{fact} J.C.\ Collins, D.E.\ Soper and G.\ Sterman, in {\it Perturbative Quantum 
Chromodynamics},
ed.\ A.H.\ Mueller (World Scientific, Singapore, 1989), p.\ 1.

\bibitem{dy1loop} G.\ Altarelli, R.K.\ Ellis and 
G.\ Martinelli, Nucl.\ Phys.\ B157 (1979) 461; B.\ Humpert and W.L.\ van Neerven, Phys.\ Lett.\ 
84B (1979) 327; J.\ Kubar-Andr\'e and F.E.\ Paige, Phys.\ Rev.\ D19 (1979) 221; 
K.\ Harada, T.\ Kaneko and N.\ Sakai, Nucl.\ Phys.\ B155 (1979) 169; 
B165 (1980) 545 (E).

\bibitem{dgamma1loop} 
P.\ Aurenche, A.\ Douiri, R.\ Baier, M.\ Fontannaz
and D.\ Schiff,
Phys.\ Lett.\ 140B (1984) 87;
H.\ Baer, J.\ Ohnemus and J.F.\ Owens,
Phys.\ Rev.\ D42 (1990) 61; P.\ Aurenche, 
P.\ Chiappetta, M.\ Fontannaz, J.P.\ Guillet, E.\ Pilon,
Nucl.\ Phys.\ B399 (1993) 34;
L.E.\ Gordon and W.\ Vogelsang, Phys.\ Rev.\ D48 (1993) 3136;
D50 (1994) 1901; M.\ Gl\'uck, L.E.\ Gordon, E.\ Reya and
W.\ Vogelsang, Phys.\ Rev.\ Lett.\ 73 (1994) 388. 

\bibitem{heavycalcs}  P.\ Nason, S.\ Dawson and R.K.\ Ellis,
Nucl.\ Phys.\ B303 (1988) 607; W.\ Beenakker, H.\ Kuijf, W.L.\ van Neerven,
and J.\ Smith Phys.\ Rev.\ D40 (1989) 54; W.\ Beenakker, W.L.\ van
Neerven, R.\ Meng, G.A.\ Schuler, and J.\ Smith, Nucl.\ Phys.\ 
B351 (1991) 507.

\bibitem{jet1loop} F.\ Aversa, P.\ Chiappetta, M.\ Greco
and J.-Ph.\ Guillet, Phys.\ Lett.\ B211 (1988) 465;
Phys.\ Rev.\ Lett.\ 65 (1990) 401; Z.\ Phys.\ C46 (1990) 253;
F.\ Aversa, P.\ Chiappetta, L.\ Gonzales, M.\ Greco
and J.-Ph.\ Guillet,
Z.\ Phys.\ C49 (1991) 459;
S.D.\ Ellis, Z.\ Kunszt and D.E.\ Soper,
Phys.\ Rev.\ D40 (1989) 2188; Phys.\ Rev.\ Lett.\  64 (1990) 2121;
{\it ibid} 69 (1992) 1496;  Z.\ Kunszt and D.E.\ Soper, Phys.\ Rev.\
D46 (1992) 192.

\bibitem{2loop} R.\ Hamberg and W.L. van Neerven, Nucl.\ Phys.\ 
{B379} (1992) 143; E.B\ Zijlstra and W.L. van Neerven, Phys.\ Lett.\ 
   {B273} (1991) 476; Nucl.\ Phys.\
   {B383} (1992) 525;   {B417} (1994) 61, 
(E) B426 (1994) 245; 
S.A.\ Larin, T.\ van Ritbergen and J.A.M.\ Vermaseren, Nucl.\ Phys.\ B427 (1994) 41;
R.\ Mertig and W.L.\ van Neerven,
Instituut-Lorentz preprint INLO-PUB-6/95,
hep-ph/9506451; 
M.\ Buza, Y.\ Matiounine, J.\ Smith, R.\ Migneron and
W.L.\ van Neerven, NIKHEF preprint NIKHEF/95-070,
hep-ph/9601302.

\bibitem{oldDY} G.\ Sterman, Nucl.\ Phys.\ B281 (1987) 310;
 S.\ Catani and L.\ Trentadue, 
Nucl.\ Phys.\ B327 (1989) 323.

\bibitem{LaenenSterman} E.\ Laenen and G.\ Sterman, in 
proceedings of {\it The Fermilab
Meeting, DPF 92}, 7th  meeting of the 
American Physical Society Division of Particles
and Fields (Batavia, IL, 1992), ed.\
C.H.\ Albright {\it et al.} (World Scientific,
Singapore, 1993), p.\ 987.

\bibitem{CSt} H.\ Contopanagos and G.\ Sterman, Nucl.\ Phys.\ B400 (1993) 211; 
B419 (1994) 77.

\bibitem{alcon}  L. Alvero and H. Contopanagos, Nucl.\ Phys.\ B436 (1995) 184;
Nucl.\ Phys.\ 
B456 (1995) 497.

\bibitem{bercon}
E.L.\ Berger and H.\ Contopanagos, Phys.\ Lett.\  B361 (1995) 115;
 Argonne preprint ANL-HEP-95-82,
hep-ph/9603326.

\bibitem{jacketal}  E.\ Laenen, J.\ Smith and W.L.\ van Neerven, 
Nucl.\ Phys.\  B369 (1992) 543;
Phys.\ Lett.\ B321 (1994) 254.

\bibitem{kidsm}
N.\ Kidonakis and J.\ Smith, Phys.\ Rev.\ D51 (1995) 6092;
Mod.\ Phys.\ Lett.\ A11 (1996) 587.

\bibitem{catanietal} S.\ Catani, M.L.\ Mangano, P.\ Nason and L.\ Trentadue,
CERN preprint CERN-TH/96-21, hep-ph/9602208; CERN-TH/96-86, hep-ph/9604351.

\bibitem{irr} D.\ Appell, P.\ Mackenzie and G.\ Sterman in {\it 
Proceedings of the Storrs Meeting}, Fourth Meeting of the Division
of Particles and Fields, Storrs, CT, August 15-18, 1988, p.\ 567;
 B.R.\ Webber, Phys.\ Lett.\ {\bf B339}, 148 (1994);  
A.V.\ Manohar and M.B.\ Wise, Phys.\ Lett.\ {\bf B344}, 407 (1995);
G.P.\ Korchemsky and G.\ Sterman, Nucl.\ Phys.\ {\bf B437}, 415 (1995);
Yu.L.\ Dokshitzer and B.R.\ Webber, 
        Phys.\ Lett.\ {\bf B352}, 451 (1995); 
   R.\ Akhoury and V.I.\ Zakharov, Phys.\ Lett.\ {\bf B357}, 646 (1995);
Nucl.\ Phys.\ B465 (1996) 295;
 P.\ Nason and M.H.\ Seymour, Nucl.\ Phys.\ {\bf B454}, 291 (1995);
M.\ Beneke and V.M.\ Braun, Nucl.\ Phys.\ {\bf B454}, 253 (1995);
Yu.L.\ Dokshitser, G.\ Marchesini and B.R.\ Webber, 
CERN-TH/95-281, hep-ph/9512336.

\bibitem{effective} K.G.\ Wilson and J.\ Kogut,
Phys.\ Rep.\ 12 (1974) 75; 
      J.\ Polchinski,
Nucl.\ Phys.\ B231 (1984) 269; H.\ Georgi,
 Ann.\ Rev.\ Nucl.\ Part.\ Sci., 43 (1993) 209.

\bibitem{BottsSt} J.\ Botts and G.\ Sterman, Nucl.\ Phys.\ B325 (1989) 62.

\bibitem{SotSt} M.\ Sotiropoulos and G.\
Sterman,  Nucl.\ Phys.\ B419 (1994) 59.

\bibitem{GK} G.P. \ Korchemsky, Phys.\ Lett.\ B325 (1994) 459.

\bibitem{KK}  I.A.\ Korchemskaya and G.P.\ Korchemsky,
Nucl.\ Phys.\ B437 (1995) 127. 

\bibitem{CLS} H. Contopanagos, E.\ Laenen, and G. Sterman, 
CERN-TH-96-75, hep-ph/9604313.

\bibitem{Sudakovff}  J.G.M.\ Gatheral, Phys.\ Lett.\ 133B (1983) 90;
J.\ Frenkel and J.C.\ Taylor, Nucl.\ Phys.\ B246 (1984) 231;
G.P.\ Korchemsky and A.V.\ Radyushkin, Phys.\ Lett.\ B171 (1986) 459;
J.C. Collins, in {\it Perturbative Quantum Chromodynamics},
ed.\ A.H.\ Mueller (World Scientific, Singapore, 1989), p.\ 573;
L.\ Magnea and G.\ Sterman, Phys.\ Rev.\ D42 (1990) 4222.

\bibitem{CT2}  S.\ Catani and L.\ Trentadue, Nucl.\ Phys.\  B353 (1991) 183.

\bibitem{mengetal} R.\ Meng, G.A.\ Schuler, J.\ Smith and W.L.\ van Neerven,
Nucl.\ Phys.\  B339 (1990) 325.

\end{thebibliography}
\end{document}